\documentclass[preprint]{ptptex}

\title{
Phenomenological formula for CKM matrix and its physical interpretation
}

\author{
Kohzo Nishida
\footnote{E-mail: EZF01671@nifty.com} 
}

\inst{
Department of Physics, Kyoto Sangyo University, Kyoto 603-8555, Japan 
}

\abst{
We propose a phenomenological formula relating the Cabibbo--Kobayashi--Masukawa matrix $V_{\mbox{\footnotesize CKM}}$ and quark masses in the form 
$(\sqrt{m_d} \, \sqrt{m_s} \, \sqrt{m_b}) \propto (\sqrt{m_u} \, \sqrt{m_c} \, \sqrt{m_t})V_{\mbox{\footnotesize CKM}}$. 
The results of the proposed formula are in agreement with experimental data. Under the constraint of the formula, 
we show that the invariant amplitude of the charged current weak interactions is maximized.
}

\begin{document}

\maketitle

Understanding the origin of the Cabibbo--Kobayashi--Masukawa (CKM) matrix\cite{1,2} is 
one of the major problems to be resolved in particle physics.
In the standard model, all components of the matrix are free parameters that 
must be adjusted experimentally. 
In the past, there have been many attempts\cite{3} to find formulas that relate the fermion masses and the CKM matrix. 
In our previous paper\cite{4}, we proposed a phenomenological formula, $(m_d \, m_s \, m_b) \propto (m_u \, m_c \, m_t)V_{\mbox{\footnotesize CKM}}$. 
In this paper we modify the previously proposed formula to fit the latest experimental data, 
and we show that the invariant amplitude of the charged current weak interactions is maximized under the constraint of the formula.

We introduce the two unit quark mass vectors as
\begin{equation}
\label{eq:10}
\mbox{\boldmath$e$}^{(u)} \equiv \frac{1}{\sqrt{m_u + m_c + m_t}} 
\left(
\begin{array}{c}
\sqrt{m_u} \\
\sqrt{m_c} \\
\sqrt{m_t}
\end{array}
\right),
  \,\,\,\,\,\,\,\,
\mbox{\boldmath$e$}^{(d)} \equiv \frac{1}{\sqrt{m_d + m_s + m_b}} 
\left(
\begin{array}{c}
\sqrt{m_d} \\
\sqrt{m_s} \\
\sqrt{m_b}
\end{array}
\right),
\end{equation}
for the up and down quark sectors.
Our basic postulate is to interpret one of the unitary matrices, $V$, satisfying
\begin{equation}
\label{eq:20}
\mbox{\boldmath$e$}^{(u)} = V \mbox{\boldmath$e$}^{(d)}
\end{equation}
as a CKM matrix.

To find such a $V$, we introduce rotation matrices for the 1--2 plane 
\begin{equation}
\label{eq:30}
R^{(q)}_{12} \equiv 
\left(
\begin{array}{ccc}
\sqrt{\frac{m^{(q)}_2}{m^{(q)}_1+m^{(q)}_2}} & -\sqrt{\frac{m^{(q)}_1}{m^{(q)}_1+m^{(q)}_2}} & 0 \\
\sqrt{\frac{m^{(q)}_1}{m^{(q)}_1+m^{(q)}_2}} &   \sqrt{\frac{m^{(q)}_2}{m^{(q)}_1+m^{(q)}_2}} & 0 \\
0 & 0 & 1
\end{array}
\right),
\,\,\,\,\,\,\,\,
q=u,d,
\end{equation}
and the 2--3 plane
\begin{equation}
\label{eq:40}
R^{(q)}_{23} \equiv 
\left(
\begin{array}{ccc}
1 & 0 & 0 \\
0 & \sqrt{\frac{m^{(q)}_3}{m^{(q)}_1+m^{(q)}_2+m^{(q)}_3}} & -\sqrt{\frac{m^{(q)}_1+ m^{(q)}_2}{m^{(q)}_1+m^{(q)}_2+m^{(q)}_3}} \\
0 & \sqrt{\frac{m^{(q)}_1+m^{(q)}_2}{m^{(q)}_1+m^{(q)}_2+m^{(q)}_3}} & \sqrt{\frac{m^{(q)}_3}{m^{(q)}_1+m^{(q)}_2+m^{(q)}_3}} \\
\end{array}
\right),
\,\,\,\,\,\,\,\,
q=u,d,
\end{equation} 
where $m^{(u)}_1 = m_u$, $m^{(u)}_2 = m_c$, $m^{(u)}_3 = m_t$, 
$m^{(d)}_1 = m_d$, $m^{(d)}_2 = m_s$, $m^{(d)}_3 = m_b$.
Note that we can always,  by the rotation matrices, convert $\mbox{\boldmath$e$}^{(u,d)}$ to the form 
\begin{equation}
\label{eq:50}
R_{23}^{(u)}R_{12}^{(u)}\mbox{\boldmath$e$}^{(u)} =  R_{23}^{(d)}R_{12}^{(d)} \mbox{\boldmath$e$}^{(d)}
= 
\left(
\begin{array}{c}
0 \\
0 \\
1
\end{array}
\right).
\end{equation}
Equation (\ref{eq:50}) can be generalized to 
\begin{equation}
\label{eq:80}
R_{23}^{(u)}R_{12}^{(u)}\mbox{\boldmath$e$}^{(u)} = U_{12} R_{23}^{(d)}R_{12}^{(d)} \mbox{\boldmath$e$}^{(d)}
= 
\left(
\begin{array}{c}
0 \\
0 \\
1
\end{array}
\right),
\end{equation}
where $U_{12}$ is the unitary matrix 
\begin{equation}
\label{eq:60}
U_{12}(\phi,\delta_1,\delta_2,\delta_3) \equiv 
\left(
\begin{array}{ccc}
e^{i\delta_1}\cos \phi & -e^{i\delta_3}\sin \phi & 0 \\
e^{i\delta_4}\sin \phi &e^{i\delta_2}\cos \phi & 0 \\
0 & 0 & 1
\end{array}
\right),
  \,\,\,\,\,\,\,\,
\delta_1 + \delta_2 = \delta_3 + \delta_4,
\end{equation}
which satisfies
\begin{equation}
\label{eq:70}
U_{12}
\left(
\begin{array}{c}
0 \\
0 \\
1
\end{array}
\right)
=
\left(
\begin{array}{c}
0 \\
0 \\
1
\end{array}
\right).
\end{equation}
From  (\ref{eq:80}), we obtain
\begin{equation}
\label{eq:90}
\mbox{\boldmath$e$}^{(u)} = (R_{23}^{(u)} R_{12}^{(u)})^{\mbox{\footnotesize T}} U_{12} R_{23}^{(d)}R_{23}^{(d)}\mbox{\boldmath$e$}^{(d)}.
\end{equation}
Hence, we may generally write $V$ in the form
\begin{equation}
\label{eq:100}
V = (R_{23}^{(u)} R_{12}^{(u)})^{\mbox{\footnotesize T}} U_{12}(\phi,\delta_1,\delta_2,\delta_3) R_{23}^{(d)}R_{23}^{(d)},
\end{equation}
with the indefinite parameters $\phi,\delta_1,\delta_2$, and $\delta_3$.

Recently, the averages of quark masses and the absolute values of CKM matrix elements have been estimated as follows\cite{5}:
$m_u = 2.2^{+0.5}_{-0.4}$MeV, $m_d = 4.7^{+0.5}_{-0.4}$MeV, $m_s = 96^{+8}_{-4}$MeV,
$m_c = 1.28\pm 0.03$GeV, $m_b = 4.18^{+0.04}_{-0.03}$GeV, $m_t = 173.1\pm 0.6$GeV,
\begin{equation}
\label{eq:110}
V_{\mbox{\footnotesize CKM}} =
\left(
\begin{array}{ccc}
0.97434^{+0.00011}_{-0.00012} & 0.22506 \pm 0.00050 & 0.00357 \pm 0.00015 \\
0.22492 \pm 0.00050 & 0.97351 \pm 0.00013 & 0.0411 \pm 0.0013 \\
0.00875^{+0.00032}_{-0.00033} & 0.0403 \pm 0.0013 & 0.99915 \pm 0.00005 \\
\end{array}
\right),
\end{equation}
and the Jarlskog invariant is $J=(3.04^{+0.21}_{-0.20})\times 10^{-5}$.

When we use the values
$m_u = 2.2$MeV, $m_d = 4.7$MeV, $m_s = 96$MeV,
$m_c = 1.28$GeV, $m_b = 4.2$GeV, $m_t = 170$GeV,
$\phi = 0.0507$, and $\delta_1 = \delta_2 = \delta_3 = 0$,
 (\ref{eq:100}) is calculated as
\begin{equation}
\label{eq:120}
V = 
\left(
\begin{array}{ccc}
0.9744 & -0.2246 & 0.004983 \\
0.2244 & 0.9722 & -0.06701 \\
0.01020 & 0.06642 & 0.9977 \\
\end{array}
\right), 
\end{equation}
where we choose $\phi = 0.0507$ by using the least-squares method minimizing 
\begin{equation}
\left. \chi^2(\phi) \right|_{\delta_1 = \delta_2 = \delta_3 = 0} = \sum^3_{i,j = 1}\{(V_{\mbox{\footnotesize CKM}})_{ij}\mbox{ of (\ref{eq:110})}
 - |V_{ij}(\phi)|\mbox{ of (\ref{eq:100})} \}^2.
\end{equation}
The absolute values of  (\ref{eq:120}) are in good agreement with the experimental data obtained in  (\ref{eq:110}).

For $m_u = 2.2$MeV, $m_d = 4.7$MeV, $m_s = 96$MeV,
$m_c = 1.28$GeV, $m_b = 4.2$GeV, $m_t = 170$GeV,
$\phi = 0.0507$, and $\delta_1 = \delta_2 = \delta_3 \equiv \delta$,
$\chi^2(\delta)|_{\phi=0.0507}$ has a minimum value at $\delta=-0.025$. 
In the case, (\ref{eq:100}) is calculated as
\begin{equation}
\label{eq:121}
|V| = 
\left(
\begin{array}{ccc}
0.9744 & 0.2246 & 0.004982 \\
0.2244 & 0.9722 & 0.06704 \\
0.01019 & 0.06637 & 0.9977 \\
\end{array}
\right),
\end{equation}
and the Jarlskog invariant $J = \mbox{Im}(V_{23}V_{12}V_{22}^{\dagger}V_{13}^{\dagger})$ is calculated as
\begin{equation}
J=-1.056\times 10^{-6}.
\end{equation}
$J$ is somewhat small in comparison  with the present data.
$J$ can take the value $3.038\times 10^{-5}$ at $\delta=-1.139$,
but the matrix  (\ref{eq:100}) calculated by the parameter set,
\begin{equation}
\label{eq:122}
|V| = 
\left(
\begin{array}{ccc}
0.9744 & 0.2249 & 0.002924 \\
0.2227 & 0.9647 & 0.1167 \\
0.008946 & 0.02014 & 0.4252 \\
\end{array}
\right),
\end{equation}
does not agree with the experimental data of $V_{33}$.

Next,  we discuss the physical meaning of the formula (\ref{eq:20}).
For a free particle, we can seek four-momentum eigensolutions of Dirac's equation of the form
\begin{equation}
\label{eq:130}
\psi = u(\mbox{\boldmath$p$})e^{-ip\cdot x},
\end{equation} 
where $u(\mbox{\boldmath$p$})$ is a four-component spinor independent of $x$.
Substituting this equation into Dirac's equation, we have
\begin{equation}
\label{eq:140}
(p\!\!\!/-m)u = 0.
\end{equation}
For $\mbox{\boldmath$p$} \neq 0$,
the positive-energy four-spinor solutions of Dirac's equation are
\begin{equation}
\label{eq:150}
u^{(s)} = N
\left(
\begin{array}{c}
\chi^{(s)} \\
\frac{\mbox{\boldmath$\sigma \cdot p$}}{\mbox{$E+m$}}\chi^{(s)} 
\end{array}
\right),
\,\,\,\,\,\,\,\,
s = 1,2,
\end{equation}
where 
\begin{equation}
\label{eq:160}
\chi^{(1)} = 
\left(
\begin{array}{c}
1 \\
0 
\end{array}
\right),
\,\,\,\,\,\,\,\,
\chi^{(2)} = 
\left(
\begin{array}{c}
0 \\
1 
\end{array}
\right).
\end{equation}
If we choose the covariant normalization in which we have 
$2E$ particles per unit volume,
the normalization constant $N$ is
\begin{equation}
\label{eq:170}
N = \sqrt{E+m}.
\end{equation}
For $|\mbox{\boldmath$p$}| \rightarrow 0$,  (\ref{eq:150}) becomes\cite{6} 
\begin{equation}
\label{eq:180}
u^{(s)} = \sqrt{2m}
\left(
\begin{array}{c}
\chi^{(s)} \\
0
\end{array}
\right).
\end{equation}
Note that $u$ is proportional to the square root of the mass.

The invariant amplitude ${\cal M}$ including weak quark currents $ (J_\mu)_q$ as $d,s,b \rightarrow u,c,t$ 
 is written as follows:
\begin{equation}
\label{eq:190}
{\cal M} = \frac{4G}{\sqrt{2}}(J^{\mu})_F (J_\mu)_q,
\end{equation}
where
\begin{equation}
\label{eq:200}
(J_\mu)_q = 
(\bar{u}_u \,\, \bar{u}_c \,\, \bar{u}_t)  \gamma_{\mu} \frac{1}{2}(1-\gamma^5) V_{\mbox{\footnotesize CKM}} 
\left(
\begin{array}{c}
u_d \\
u_s \\
u_b
\end{array}
\right).
\end{equation}
By substituting  (\ref{eq:180}) into  (\ref{eq:190}), for $|\mbox{\boldmath$p$}| \rightarrow 0$, we derive
\begin{eqnarray}
\label{eq:210}
{\cal M} 
&\sim& (J_\mu)_q \nonumber \\ 
&=& 
(\chi^{(s)} \,\, 0)  \gamma_{\mu} (1-\gamma^5) 
\left(
\begin{array}{c}
\chi^{(s)} \\
0
\end{array}
\right)
(\sqrt{m_u} \,\, \sqrt{m_c} \,\, \sqrt{m_t}) V_{\mbox{\footnotesize CKM}}  
\left(
\begin{array}{c}
\sqrt{m_d} \\
\sqrt{m_s} \\
\sqrt{m_b}
\end{array}
\right) \nonumber \\
&=&
(\chi^{(s)} \,\, 0)  \gamma_{\mu} (1-\gamma^5) 
\left(
\begin{array}{c}
\chi^{(s)} \\
0
\end{array}
\right) \nonumber \\
&&
\times \sqrt{m_u+m_c+m_t}\sqrt{m_d+m_s+m_b} \, \mbox{\boldmath$e$}^{(u)} \cdot V_{\mbox{\footnotesize CKM}} \mbox{\boldmath$e$}^{(d)} \nonumber \\
&=&
(\chi^{(s)} \,\, 0)  \gamma_{\mu} (1-\gamma^5) 
\left(
\begin{array}{c}
\chi^{(s)} \\
0
\end{array}
\right) 
\sqrt{m_u+m_c+m_t}\sqrt{m_d+m_s+m_b} \, \cos \theta,
\end{eqnarray}
where $\theta$ is the angle between the unit vectors $\mbox{\boldmath$e$}^{(u)}$ and $V_{\mbox{\footnotesize CKM}} \mbox{\boldmath$e$}^{(d)}$.
Hence, ${\cal M}$ has a maximum value at $\theta = 0$, that is, when $\mbox{\boldmath$e$}^{(u)} = V_{\mbox{\footnotesize CKM}} \mbox{\boldmath$e$}^{(d)}$.

For $|\mbox{\boldmath$p$}| \rightarrow 0$,
the electroweak interaction term in the Lagrangian on substituting  (\ref{eq:180}) becomes
\begin{eqnarray}
\label{eq:220}
{\cal L} 
&=&
\cdots -i\frac{g}{\sqrt{2}} (J^\mu)_q W^+_{\mu} \nonumber \\
&=& \cdots 
-i\frac{g}{\sqrt{2}}  (\chi^{(s)} \,\, 0)  \gamma^{\mu} (1-\gamma^5) 
\left(
\begin{array}{c}
\chi^{(s)} \\
0
\end{array}
\right)
(\sqrt{m_u} \,\, \sqrt{m_c} \,\, \sqrt{m_t}) V_{\mbox{\footnotesize CKM}}  
\left(
\begin{array}{c}
\sqrt{m_d} \\
\sqrt{m_s} \\
\sqrt{m_b}
\end{array}
\right)
  W^+_{\mu} \nonumber \\
&=& \cdots 
-i\frac{g}{\sqrt{2}}  (\chi^{(s)} \,\, 0)  \gamma^{\mu} (1-\gamma^5) 
\left(
\begin{array}{c}
\chi^{(s)} \\
0
\end{array}
\right)  W^+_{\mu} \nonumber \\
&&\times \sqrt{m_u+m_c+m_t}\sqrt{m_d+m_s+m_b}\cos \theta.
\end{eqnarray}
If we assume that the quark mixing angles are dynamic parameters, 
the variation of the Lagrangian is 
\begin{eqnarray}
\label{eq:240}
\delta {\cal L} 
&=& \cdots 
-i\frac{g}{\sqrt{2}}  (\chi^{(s)} \,\, 0)  \gamma^{\mu} (1-\gamma^5) 
\left(
\begin{array}{c}
\chi^{(s)} \\
0
\end{array}
\right)  W^+_{\mu} \nonumber \\
&&\times \sqrt{m_u+m_c+m_t}\sqrt{m_d+m_s+m_b}\sin \theta \delta \theta.
\end{eqnarray}
When we require $\delta {\cal L} = 0$,
we obtain $\theta = 0$, that is, $\mbox{\boldmath$e$}^{(u)} = V_{\mbox{\footnotesize CKM}} \mbox{\boldmath$e$}^{(d)}$.

Thus, we conclude that the origin of the quark mixing is the most likely configuration of scattering.
The conclusion predicts that the value of the CKM matrix elements has momentum dependence because the mixing angle to maximize ${\cal M}$ of (\ref{eq:190}) depends on the momentum of each particle.
Then the formula $\mbox{\boldmath$e$}^{(u)} = V \mbox{\boldmath$e$}^{(d)}$ 
is a zeroth approximation.
In the first approximation, we will need to replace the components $\sqrt{m^{(q)}_i}$ 
of $\mbox{\boldmath$e$}^{(q)}$ by $\sqrt{E^{(q)}_i + m^{(q)}_i}$.


\begin{thebibliography}{9}

\bibitem{1}
N. Cabibbo, Phys. Rev. Lett. {\bf 10}, 531 (1963).

\bibitem{2}
M. Kobayashi and T. Maskawa, Prog. Theor. Phys. {\bf 49}, 652 (1973).

\bibitem{3}
H.Fritzsch, Phys. Lett.  {\bf B73}, 317 (1978); Nucl. Phys.  B {\bf 155}, 189 (1979).

\bibitem{4}
K. Nishida and I. S. Sogami, Prog. Theor. Phys. {\bf 96}, 857 (1996)[arXiv:hep-ph/9608353].

\bibitem{5}
C. Patrignani et al. [Particle Data Group], Chin. Phys. C {\bf 40}, 100001 (2016) and 2017 update.

\bibitem{6}
F. Halzen and A. D. Martin, {\it Quarks and Leptons} (John Wiley \& Sons, New York, 1984).

\end{thebibliography}
\end{document}